\documentclass{elsart}
\usepackage{graphicx}
\usepackage{epsfig}
\usepackage{amssymb}
\usepackage{lineno}
\usepackage{rotating,multirow}
\usepackage{arydshln}
\usepackage{appendix}
\usepackage[latin1]{inputenc}
\pdfoutput=1
\newcommand{\degs}{\mbox{\(^\circ \)}}

\begin{document}
\begin{frontmatter}

\title{Correlation of the highest-energy cosmic rays\\
with the positions of \\
nearby active galactic nuclei}

% This is a fragment that can be inserted into a LaTeX document:
\par\noindent
{\bf The Pierre Auger Collaboration} \\
\author{\small J.~Abraham$^{14}$}, 
\author{\small P.~Abreu$^{66}$}, 
\author{\small M.~Aglietta$^{52}$}, 
\author{\small C.~Aguirre$^{17}$}, 
\author{\small D.~Allard$^{32}$}, 
\author{\small I.~Allekotte$^{7}$}, 
\author{\small J.~Allen$^{85}$}, 
\author{\small P.~Allison$^{87}$}, 
\author{\small J.~Alvarez-Mu\~{n}iz$^{73}$}, 
\author{\small M.~Ambrosio$^{55}$}, 
% ...

\par\noindent
{\tiny
% ...
$^{7}$ Centro At\'{o}mico Bariloche, Comision Nacional de Energ\'{\i}a 
At\'{o}mica and Instituto Balseiro (CNEA-UNC), San Carlos de 
Bariloche, Argentina \\
% ...
$^{14}$ Universidad Tecnol\'{o}gica Nacional, FR-Mendoza, Argentina
 \\
% ...
$^{17}$ Universidad Catolica de Bolivia, La Paz, Bolivia \\
% ...
$^{32}$ Laboratoire AstroParticule et Cosmologie, Universit\'{e} 
Paris 7, IN2P3/CNRS, Paris, France \\
% ...
$^{52}$ Istituto di Fisica dello Spazio Interplanetario (INAF),
 Universit\`{a} di Torino and Sezione INFN, Torino, Italy \\
% ...
$^{55}$ Sezione INFN di Napoli, Napoli, Italy \\
% ...
$^{66}$ LIP and Instituto Superior T\'{e}cnico, Lisboa, Portugal \\
% ...
$^{73}$ Universidad de Santiago de Compostela, Spain \\
% ...
$^{85}$ New York University, New York, NY, USA \\
% ...
$^{87}$ Ohio State University, Columbus, OH, USA \\
% ...
}
% last updated:	1/28/2008 

\begin{abstract}
Data collected by the Pierre Auger Observatory provide evidence for
anisotropy in the arrival directions of the cosmic rays with the
highest energies, which are correlated with the positions of
relatively nearby active galactic nuclei (AGN) \cite{science}. The
correlation has maximum significance for cosmic rays with energy
greater than $\sim\ 6\times 10^{19}$~eV and AGN at a distance less
than $\sim\ 75$~Mpc. We have confirmed the anisotropy at a confidence
level of more than 99\% through a test with parameters specified {\em
a priori}, using an independent data set.  The observed correlation is
compatible with the hypothesis that cosmic rays with the highest
energies originate from extra-galactic sources close enough so that
their flux is not significantly attenuated by interaction with the
cosmic background radiation (the Greisen-Zatsepin-Kuz'min effect).
The angular scale of the correlation observed is a few degrees, which
suggests a predominantly light composition unless the magnetic fields
are very weak outside the thin disk of our galaxy. Our present data do
not identify AGN as the sources of cosmic rays unambiguously, and
other candidate sources which are distributed as nearby AGN are not
ruled out.  We discuss the prospect of unequivocal identification of
individual sources of the highest-energy cosmic rays within a few
years of continued operation of the Pierre Auger Observatory.
\end{abstract} 

\end{frontmatter}
\section{Introduction}

The identification of the sources of the cosmic rays with the highest
energies so far detected has been a great challenge ever since the
first event with energy around $10^{20}$~eV was
reported~\cite{linsley}.  If the highest-energy cosmic rays are
predominantly protons and nuclei, only sources which are less than
about 200~Mpc from Earth could contribute significantly to the
observed flux above $6\times 10^{19}$~eV. Protons with higher energies
interact with cosmic microwave background photons to produce
pions~\cite{GZK-1,GZK-2}, which leads to a significant attenuation of
their flux from more distant sources. The energy of light nuclei is
damped over an even shorter length scale due to photo-disintegration
processes~\cite{heavy-1,heavy-2}.  If the relatively nearby sources are not
uniformly distributed then we expect that the arrival directions of
the most energetic cosmic rays should be anisotropic, as long as
deflections imprinted by intervening magnetic fields upon their
trajectories are small enough that they point back to their place of
origin.

The Pierre Auger Observatory~\cite{nim}, has been operating in
Argentina and taking data in a stable mode since January 2004.  The
large exposure of the surface detectors (SD), combined with accurate
energy and arrival direction measurements, calibrated and verified
from the hybrid operation with fluorescence detectors (FD), provides
an opportunity to find the clues that could lead to 
an understanding of the origin of the highest-energy cosmic rays.

AGN have long been considered possible sites for energetic particle
production, where protons and heavier nuclei could be accelerated up
to the highest energies measured so far~\cite{Ginzburg,Hillas}.
Windows of a few degrees around each known AGN lying within 100 Mpc
cover a significant fraction -- but not most -- of the sky. We were
therefore motivated to search for an excess, as compared to
expectations for an isotropic flux, of cosmic rays with arrival
directions close to AGN.  The angular size of the search window  should
not be limited to the instrumental angular resolution, since
correlation could exist on larger scales due to magnetic deflections,
the precise amount of which is unknown.  Arrival directions of cosmic
rays are reconstructed by the SD array with an angular accuracy better
than $1^\circ$ above $10^{19}$~eV~\cite{ave}.

We have recently reported \cite{science} the observation of a
correlation between the arrival directions of the cosmic rays with
highest energies measured by the Pierre Auger Observatory and the
positions of nearby AGN from the $12^{th}$ edition of the catalogue of
quasars and active nuclei by V\'eron-Cetty and V\'eron~\cite{VC06}. In
this article we provide more details about the methods used to
demonstrate anisotropy based on this correlation, and further analyse
its properties and implications.

\section{Evidence for anisotropy and correlation with AGN} 

\subsection{Data set}
\label{s:dataset}

The southern site of the Pierre Auger Observatory \cite{nim} is
located in Malarg\"ue, Argentina, at latitude $35.2^\circ$ S,
longitude $69.5^\circ$ W, and mean altitude 1400 meters above sea
level.  The data set analysed here consists of
events recorded by the Pierre Auger Observatory from 1 January 2004
to 31 August 2007.  During this time, the size of the Observatory
increased from 154 to 1388 surface detector stations. We consider 
events with reconstructed energies above 40 EeV
(1 EeV = $10^{18}$~eV) and zenith angles smaller than $60^\circ$. The
quality cut implemented in the present analysis requires that at least
five active nearest neighbours surround the station with the highest
signal when the event was recorded, and that the reconstructed shower
core be inside an active equilateral triangle of detectors.

The event direction is determined by a fit of the arrival times of the shower front at the SD.
The precision achieved in the arrival direction depends on the clock
resolution of each detector and on the fluctuations in the time of arrival of the first
particle~\cite{timevariance}. The angular resolution is defined as the angular
aperture around the arrival directions of cosmic rays within which 68\% of
the showers are reconstructed. This resolution has been verified experimentally~\cite{carla,ave}.
Almost all events with energies above 10~EeV  trigger at least 6 surface stations and have an angular resolution
better than $1^\circ$~\cite{carla,ave}.

The energy of each event is determined in a two-step procedure. The shower size $S$, at a
reference distance and zenith angle, is calculated from the signal detected in each surface station and then converted to energy using a linear calibration curve based on the fluorescence telescope measurements~\cite{ro07}. The uncertainty in $S$ resulting from the adjustment of the shower size, the conversion to a reference angle, the fluctuations from shower-to-shower and the calibration curve amounts  to about 18\%.
The absolute energy scale is given by the fluorescence measurements and has a systematic uncertainty of 22\%~\cite{bruce}.
There is an additional uncertainty in the energy scale for the set of high energy
events used in the present analysis due to the relatively low statistics available for calibration in this energy range.

\subsection{Exposure}
The integrated exposure for the event selection described in the previous section 
amounts to  9,000~km$^2$~sr~yr. Note that analyses involving a flux calculation, such 
as the estimate of the cosmic-ray spectrum~\cite{ro07}, use stricter 
selection criteria which would amount to an exposure of about 
7000~km$^2$~sr~yr for the same data period. 

The surface detector array has full acceptance for events with energy above 3~EeV
\cite{allard}. Above this energy the detection efficiency is larger than 99\% and it is 
nearly independent  of the direction of the shower axis  defined by the zenith angle ($\theta$) 
with respect to the local vertical and azimuth ($\phi$) with respect to the South. 
Thus, above that energy the instantaneous instrument aperture as a 
function of zenith angle is given by~:

\begin{equation}
A(t) = n(t)a_0\cos{\theta}\d\Omega \d t 
\end{equation}

where $a_0\cos{\theta}$ is the surface of a unitary cell under the incidence zenith angle $\theta$ 
and $n(t)$ is the number of active such cells as a function of time. The number $n(t)$ is recorded every 
second  by the trigger  system of the Observatory and reflects the array growth as well as the dead 
period of each detector. Such recording allows for a precise knowledge of our aperture at any 
moment in time.

The instrument exposure above a certain energy $E$ may be further affected by
the conversion of the measured signal at ground to energy (this dependence is not 
included in Eq.~1 above).  For a given energy $E$ the ground signal vary depending
on the atmospheric conditions (e.g. through the variations of the Molière radius)~\cite{bleve}. 
If the signal to energy conversion does not correct  for these small variations, of order a few \%, 
as it is the case in our analysis, the aperture above a certain uncorrected energy will 
depend on the atmospheric conditions.

Over the period from 1 January 2004 to 31 August 2007 the integration of the time dependences 
from the array growth and dead time together with the atmospheric variations  introduce  a 
modulation of the exposure as a  function of celestial right ascension (RA) of less than 1\%. 
For the purpose of our analysis, 
where the total number  of events considered is less than 100, such modulation is negligible 
and the resulting RA dependence can be safely ignored. 

Hence our exposure only depends on the celestial declination $\delta$  and can be
derived from the relation $\sin\delta=\cos\theta\sin \lambda - \sin\theta\cos \lambda \cos\phi$, 
where $\lambda$ is the latitude of the Observatory.  

\subsection{Search method}
\label{s:method}

We denote by $p$ the probability that an individual event from an 
isotropic flux has, by 
chance, an arrival direction closer than some
particular angular distance $\psi$ from any member of a collection of
candidate point sources. $p$ is the
exposure-weighted fraction of the sky accessible to observation by
the Pierre Auger Observatory which is covered 
by windows of radius $\psi$ centred on the selected sources. 

The probability $P$ that $k$ or more out of a total of $N$
events from an isotropic flux are correlated by chance with the
selected objects at the chosen angular scale
is given by the cumulative binomial distribution:

\begin{equation}
P=\sum_{j=k}^N \left(\matrix{N\cr j}\right) p^{\/j}(1-p)^{N-j}~.
\label{P}
\end{equation}

For this analysis we consider the correlation between cosmic rays and
AGN in the 12$^{th}$ edition of the catalogue of quasars and active
nuclei by V\'eron-Cetty and V\'eron~\cite{VC06} (V-C). This catalogue can not
be claimed to contain all existing AGN, nor to be an unbiased
statistical sample of them. It, however, contains the results of a thorough survey of all
such objects in the literature. This catalogue contains 85,221 quasars,
1,122 BL Lac objects and 21,737 active galaxies. Among these objects,
694 have redshift $z\le 0.024$, a value corresponding to a distance
smaller than approximately 100 Mpc.\footnote{For a redshift $z$ small compared to 1,
the distance to an object is approximately $42~{\rm Mpc} \times (z/0.01)$ for
a Hubble constant $H_0=71~{\rm km~s^{-1}Mpc^{-1}}$.} At distances
greater than 100~Mpc the catalogue becomes increasingly incomplete and
inhomogeneous. The V-C catalogue is also particularly
incomplete around the galactic plane. This is not an obstacle to
demonstrating the existence of anisotropy but may affect our ability
to identify the cosmic-ray sources unambiguously.

We compute the degree of correlation as a function of three
parameters: the maximum AGN redshift $z_{max}$, the maximum angular
separation $\psi$, and the lower threshold energy for cosmic
rays $E_{th}$.  Our scan in angular separation $\psi$ is constrained
by the angular resolution of the SD \cite{ave} at the low end (we use
$\psi_{min}=1^\circ$) and by the increase in the individual
probability $p$ at the high end; large $\psi$ push the value of $p$
toward unity, rendering searches for correlation above isotropic
expectations meaningless. Illustrative values are $p=0.27$
and $p=0.6$ for maximum angular distance $\psi=3^\circ$ and
$\psi=6^\circ$, respectively, in the case of maximum AGN redshift
$z_{max}=0.024$.  Our scan in energies is motivated by the assumption
that the highest-energy cosmic rays are those that are least
deflected by intervening magnetic fields, and that they have a smaller
probability to arrive from very distant sources due to the GZK effect
\cite{GZK-1,GZK-2}.  The scan in energy threshold is carried out starting with the
event with the highest energy and adding, one by one, events with 
successively lower energy. 

We scan with the method described above to find the minimum value of 
$P$, given in Eq.~\ref{P}. Note however that $P_{min}$ is not the
chance probability that the observed arrival directions are isotropically 
distributed. An estimate of the chance probability must incorporate the
effect of the scan performed upon the data.  To do so, we build simulated 
sets, 
each having the same number of events as in the data set,
drawn from an 
isotropic flux in proportion to the relative exposure of the Observatory.
The chance probability is estimated from the fraction of simulated isotropic 
sets that have, anywhere in the parameter space and 
under the same scan, equal or smaller values of $P_{min}$ than the minimum 
found in the data \cite{FandW}.   The result can only be considered an 
estimate of the chance probability, since it depends 
somewhat on the choice of the range for the scan parameters, and does not account for the
possibility of dilution due to different scan methods or for scans
against different sets of astronomical objects.

\subsection{Exploratory scan and anisotropy confirmation}

An exploratory search for correlation between cosmic rays and AGN was
conducted according to the method described in Section~\ref{s:method}
using data collected from 1~January  2004 to 27~May  2006. 
This search yielded a minimum probability $P_{min}$ for the parameter 
set: $z_{max}= 0.018$
($D_{max}=75$~Mpc), $E_{th}= 56$~EeV and $\psi = 3.1\degs$, with
12 events among 15 correlated with at least one of the selected AGN.
For this parameter set, the chance correlation is $p=0.21$. Only 
3.2 events
were expected to correlate by chance if the flux was isotropic.  

Much of the discussion regarding past evidence for possible anisotropy in the distribution of the arrival directions of ultra-high energy cosmic rays has been centred on the issue of the impact of trial factors on the statistical significance of any potential signal. 
An accurate measure of the statistical significance of some previous reports of anisotropy ~\cite{agasagv,sugargv,agasaauto,claimbllacs}, could not be achieved due to the posterior nature of the analyses involved.  It is only from subsequent observations that those claims could be quantitatively evaluated~\cite{augergv,hiresauto,hiresbllacs,augerauto,augerbllacs}.

Therefore, to avoid the negative impact of trial factors in {\em 
a posteriori} anisotropy searches, 
the Pierre Auger collaboration
decided that any potentially interesting anisotropy signal should be tested
on an independent data 
set with parameters specified {\em a priori}. This method was described in~\cite{Clay}
where a particular set of parameters and sources were proposed and subsequently
tested on the first Auger data set~\cite{RevenuICRC}. 

The correlation observed in the exploratory scan motivated the 
construction of a specific test to 
reject or accept the isotropy hypothesis
with parameters specified {\em a priori} on an independent data set,
using exactly the same reconstruction algorithms, energy calibration
and quality cuts for event selection as in the exploratory scan.  All
details of the prescribed test were documented and archived in an
internal note. 

The test null hypothesis is isotropy and its statistical
characteristics are fully defined by the choice of two probabilities
known as the type I and type II errors. The type I error ($\alpha$) is
the probability of rejecting the null hypothesis incorrectly. In
our case this is the probability of declaring our independent data set 
anisotropic when it is not.  We have chosen $\alpha=1$\%.  
The type II error ($\beta$) is the probability of accepting the null hypothesis incorrectly. 
In our case this is the probability of declaring the independent data set isotropic when it is not.
We have chosen $\beta=5$\%.

The selection and correlation criteria for the events were chosen according to
the parameter set that minimised the probability in the
exploratory scan ($\psi=3.1\degs, z_{max}=0.018, E_{th}=56$~EeV).
Since we could not predict how many events would be
required to confirm the results at a statistically significant level from the exploratory
scan, we adopted a {\em running prescription} (with a pre-defined
stopping rule) for conducting a {\em sequential analysis} with
individual tests to be applied after the detection of each subsequent event
passing our selection criteria.  

If, in the sequence, one of the individual tests is satisfied, we reject the hypothesis
of isotropy with a confidence level of at least (1-$\alpha$)=99\%.
The total length of the test sequence (34 events) was determined by the requirement of detecting a minimum
correlation power of 60\%, as estimated from the statistics of the exploratory scan, within our specified
$\beta$ of 5\%.  In Table~1 we list, for a given number of events passing our selection
criteria $N$, the minimum number of events in correlation $k_{min}$ necessary to reject the null 
hypothesis (isotropy) with a  confidence level larger than 99\%, accounting for the
sequential nature and finite length of the test.
Note that for some values of $N$ (e.g., 5, 7, 11, etc.)  there exists no value of $k_{min}$ that can
satisfy the threshold probability without also having already satisfied the threshold at a lower value of $N$.

\begin{table}[t]
\begin{center}
\begin{tabular}{|l|cc|c|ccccccc|}\hline
$N$ & 4 & 6 & {\bf 8 }& 10 & 12  &...& 30 & 31 & 33 & 34\\ 
$k_{min}$ & 4 & 5 & {\bf 6} & 7 & 8 & ... & 14 & 14 & 15 &15 \\ 
\hline 
\end{tabular}
\label{t:running}
\caption{Criteria for our {\em running prescription} where $N$
corresponds to the total number of  events observed at any point 
during the {\em sequential analysis} of up to 34 events arriving with 
energy $E >$ 56 EeV. $k_{min}$ is the minimum number of events
within the angular window ($\psi=3.1\degs$), and a maximum AGN redshift
($z_{max}=0.018$) required to reject isotropy with at least a 99\%
confidence level.
This prescription applied to data
collected after 27 May 2006 was satisfied with $N=8$ and $k=6$ on 25 May
2007.}
\end{center} 
\medskip
\end{table}

\begin{figure}[t]
\centerline{\includegraphics[width=1.05\textwidth]{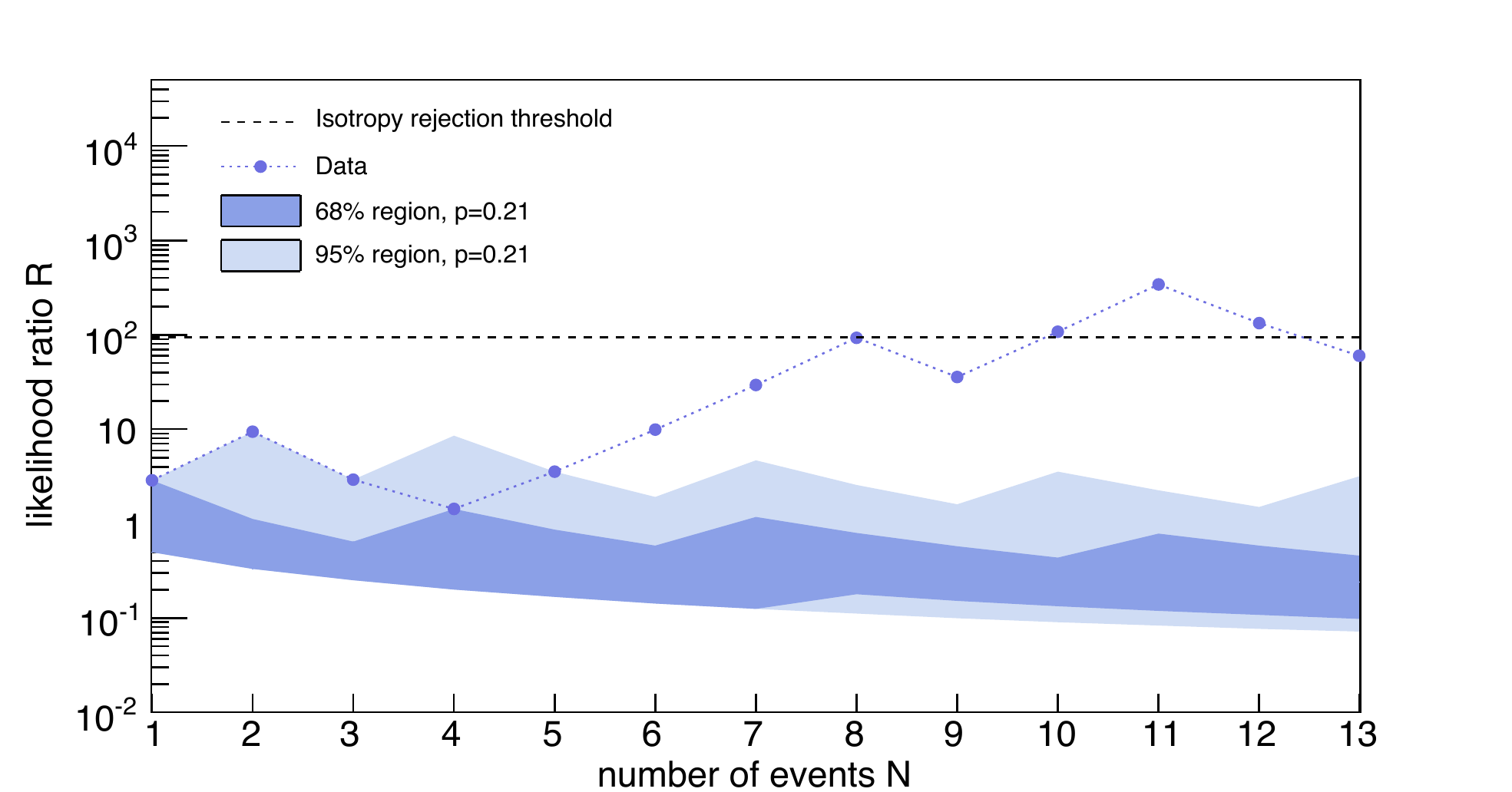}}
\caption{Likelihood ratio $R$ as a function of the number of events
observed in the prescribed test.  The null hypothesis (isotropy) was
rejected at the 99\% likelihood threshold with 10 events. Shaded
regions indicate expectations from isotropy at the 68\% and 95\% confidence 
limit.}
\label{lratio}
\end{figure}

The prescribed test was applied to data collected after 27 May 2006,
with exactly the same reconstruction algorithms, energy calibration
and quality cuts for event selection as in the exploratory scan.  On
25 May 2007, 6 out of 8 events correlated, thus satisfying
the prescription. In the independent data set collected up to 31 August 
2007
there are 13 events with energy above 56~EeV, of which 8 have arrival
directions closer than 3.1$^\circ$ from the positions of AGN less than
75 Mpc away, with 2.7 expected on average if the arrival directions were isotropic.
The probability for this single configuration to happen by chance if the
flux were isotropic (Eq.~\ref{P}) is $P=1.7\times 10^{-3}$.

Following our search protocol and based on the independent data set
alone, we reject the hypothesis of isotropy
of the arrival directions of the highest-energy cosmic rays
with at least 99\% confidence level.

An alternative standard technique in sequential analysis could also have been 
used to monitor the evolution of the correlation signal: the sequential 
likelihood ratio test~\cite{Wald:1945,Wald:1947}.  
For the sequential test of AGN correlation, the likelihood ratio $R$
is given by the relative binomial probabilities of the isotropic
(binomial parameter $p=0.21$ in our case) and anisotropic (binomial
parameter $p_1>p$) cases.  Since $p_1$ is not known, we integrate over
$p<p_1<1$ to obtain the test ratio, $R$ as defined by
Wald~\cite{Wald:1947,Stef:2007}

\begin{equation}
  R = \frac{\int_{p}^1 p_1^k(1-p_1)^{N-k}\ dp_1}{p^k(1-p)^{N-k+1}}
  \mbox{.}
  \label{eq:likelihood_ratio_marginal}
\end{equation}

The test rejects the isotropy hypothesis whenever
$R\ge(1-\beta)/\alpha$ (95 in our case) with the type I error
$\alpha=1\%$ and type II error $\beta=5$\% as previously specified.

This ratio test concluded at the rejection of the isotropy hypothesis
when 7 out of the first 10 events of our independent data set
correlated with AGN locations (see Figure~\ref{lratio}, which also
shows the subsequent evolution of the signal).

\section{The AGN correlation signal}
Having determined that an anisotropy exists according to an {\em a
priori} search over an independent subset of the Auger data, we now
consider results using the full data set (1 January 2004 - 31 August
2007) which allows us to obtain a more accurate measurement of the
correlation signal.  This data set, constructed using an updated
version of our reconstruction algorithm (see appendix A), contains 81
events with energy above 40~EeV and zenith angle smaller than
$60\degs$, which satisfy the quality criteria given in
section~\ref{s:dataset}.

\subsection{Maximum correlation parameters}
Using the method described in Section~\ref{s:method} applied to the
full data-set, we performed a scan within the range of parameters 
$1^\circ\le \psi\le 8^\circ$, $0\le z_{max}\le 0.024$ 
and $E_{th} \ge 40$~EeV. 
Catalogue-incompleteness prevents reliable exploration of higher redshifts.
The scan in maximum angular distance is performed in
steps of $0.1^\circ$, and the scan
in maximum redshift $z_{max}$ is done in steps of 0.001.

The minimum probability for the hypothesis
of isotropic arrival directions is found for the parameter set
$z_{max}= 0.017$ ($D_{max}\approx 71 $~Mpc), $\psi=3.2^\circ$, and 
$E_{th}=57$~EeV. These results are statistically consistent with the results
obtained from the earlier exploratory scan.

With these selected parameters, we find that 20 out of 27 cosmic-ray
events correlate with at least one of the 442 selected AGN (292 in the
field of view of the Observatory), while only 5.6 are expected on
average to do so if the flux were isotropic ($p=0.21$). The
respective cumulative binomial probability (Eq.~\ref{P}) of achieving
this level of correlation from an isotropic 
distribution is $P_{min}=4.6\times 10^{-9}$.  The chance 
probability that the observed correlation arose from an isotropic flux
is much larger than $P_{min}$, as already discussed in section \ref{s:method},
because a scan was performed over a large parameter space to find the minimum of $P$. 

To account for the effects of the scan we built
simulated sets each with equal number of arrival directions (81 in our
case) drawn from an isotropic flux in proportion to the relative exposure 
of the Observatory, and counted the fraction of
simulated sets which had, anywhere in the parameter space and under the
same scan, equal or smaller values of $P_{min}$ than the minimum found
in the data \cite{FandW}. 
With this procedure, we obtained smaller or equal  values of $P_{min}$ in 10$^{-5}$ of the simulated sets.

\begin{figure}
\centerline{\includegraphics[width=0.96\textwidth]{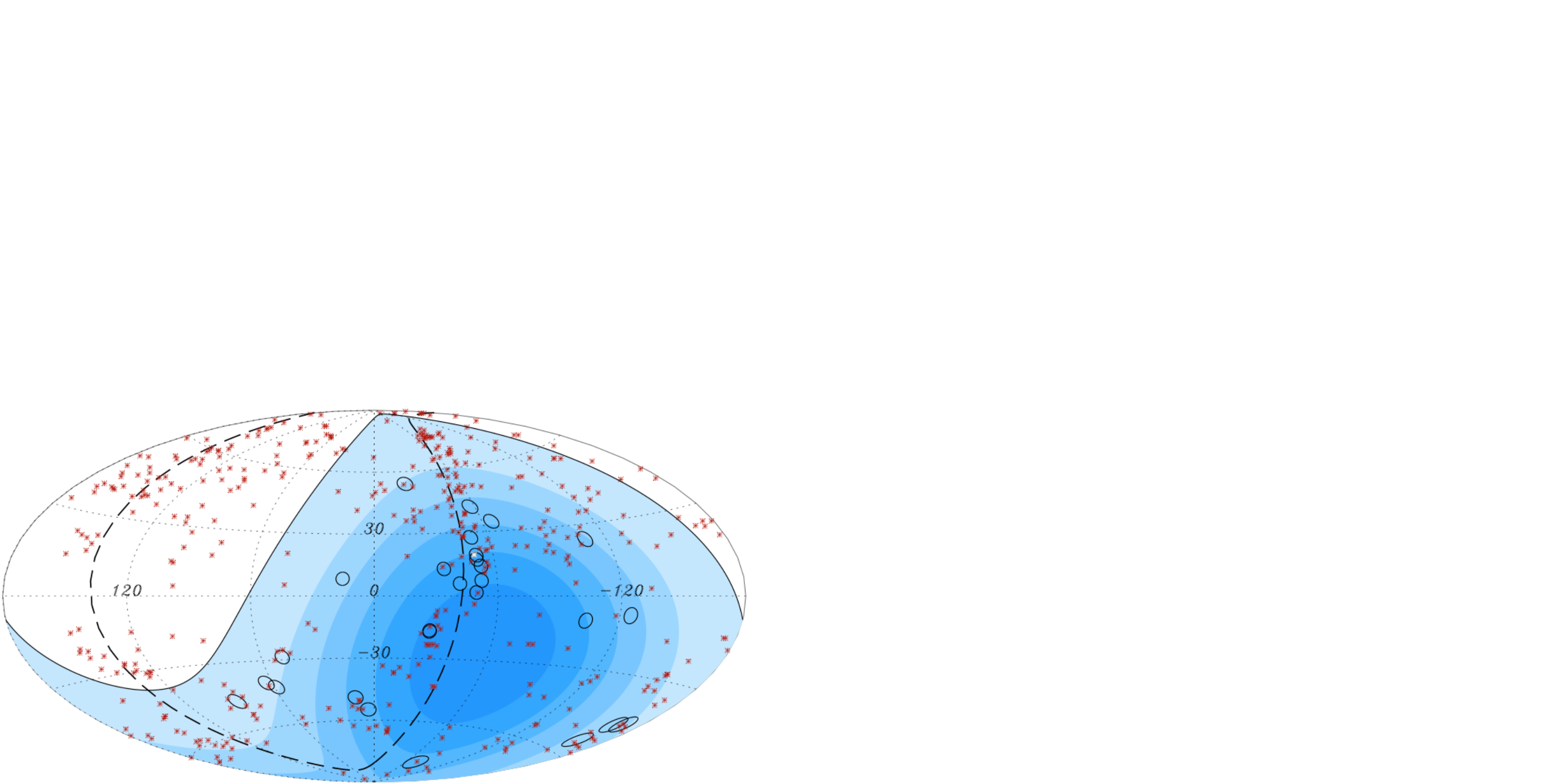}}
\caption{Aitoff projection of the celestial sphere in galactic coordinates 
with circles of $3.2^\circ$ 
centred at the arrival directions of 27 cosmic rays detected by the 
Pierre Auger Observatory
with reconstructed energies $E > 57$~EeV. The positions of the
442 AGN  (292 within the field of view of the Observatory) with redshift 
$z\le 0.017$ ($D < 71$ Mpc)  from the $12^{th}$ 
edition of the catalogue of quasars and active nuclei \cite{VC06}
are indicated by asterisks.
The solid line draws the border of the field of view for the southern 
site of the Observatory (with zenith angles smaller than $60^\circ$). The 
dashed line is, for reference,  the super-galactic plane. Darker colour 
indicates larger relative exposure. Each coloured band has equal 
integrated exposure. Centaurus A, one of the closest AGN, is marked in 
white.} \label{skymap} \end{figure}

In Figure~\ref{skymap} we present a sky map, in galactic coordinates,
with circles of radius $3.2^\circ$ around each of the arrival
directions of the 27 events with energy $E > 57$~EeV detected by the
Pierre Auger Observatory, along with asterisks at the positions of the
442 AGN with redshift $z \le 0.017$ in the V-C catalogue.  Each coloured 
band represents an equal integrated exposure which varies by about a 
factor of 3 between the lightest and darkest band. The number of AGN in 
each of those 6 bands is given in Table~\ref{t:AGNbin}. The energies
and arrival directions of the events are listed in
Appendix~\ref{EventList}.

\begin{table}[t]
\begin{center}
%\begin{tabular}{|||c|c|c|cccc|}\hline
\begin{tabular}{cccc}\hline
Declination range & Aperture fraction & Sky fraction & Number of AGN \\ \hline
$-90^\circ<\delta<-57.3^\circ$ & 1/6 & 0.08 & 25 \\ \hdashline
$-57.3<\delta<-42.3^\circ$ & 1/6 & 0.08 & 24 \\ \hdashline
$-42.3^\circ<\delta<-29.5^\circ$ & 1/6 & 0.09 & 46 \\ \hdashline
$-29.5^\circ<\delta<-16.8^\circ$ & 1/6 & 0.10 & 27 \\  \hdashline
$-16.8^\circ<\delta<-2.4^\circ$ & 1/6 & 0.12 & 63 \\  \hdashline
$-2.4^\circ<\delta<24.8^\circ$ & 1/6 & 0.23 & 107 \\  \hline
$24.8^\circ<\delta<90^\circ$ & 0 & 0.29 & 150 \\ 
%Declination range & $[-90^\circ,-57.3^\circ]$ & $[-57.3,-42.3^\circ$] & $[-42.3^\circ,-29.5^\circ$]& $[-29.5^\circ,-16.8^\circ$]& $-16.8^\circ<\delta<-2.4^\circ$  & $-2.4^\circ<\delta<24.75^\circ$ & $\delta>24.75^\circ$\\ 
%Aperture fraction & 1/6 & 1/6 & 1/6 & 1/6 & 1/6 & 1/6 & 0 \\
%Sky fraction & 0.08 & 0.09 & 0.11 & 0.12 & 0.23 & 0.37 \\
%Number of AGN & 18 & 27 & 15 & 22 & 56 & 130 & 174 \\
\hline 
\end{tabular}
\vspace*{0.2cm}
\caption{\label{t:AGNbin}Number of AGN with $z<=0.017$ in each of the exposure bands indicated in 
Fig.~\ref{skymap}. Each of the top 6 bands represent 1/6 of the total exposure, the corresponding fraction 
of the whole sky is also indicated. The last declination band  represents the part of the sky 
outside the field of view of Auger for zenith angles $\theta<60^\circ$. 
}
\end{center} 
\medskip
\end{table}

\subsection{Properties of the correlation signal}
\begin{figure}
\centerline{\includegraphics[width=\textwidth]{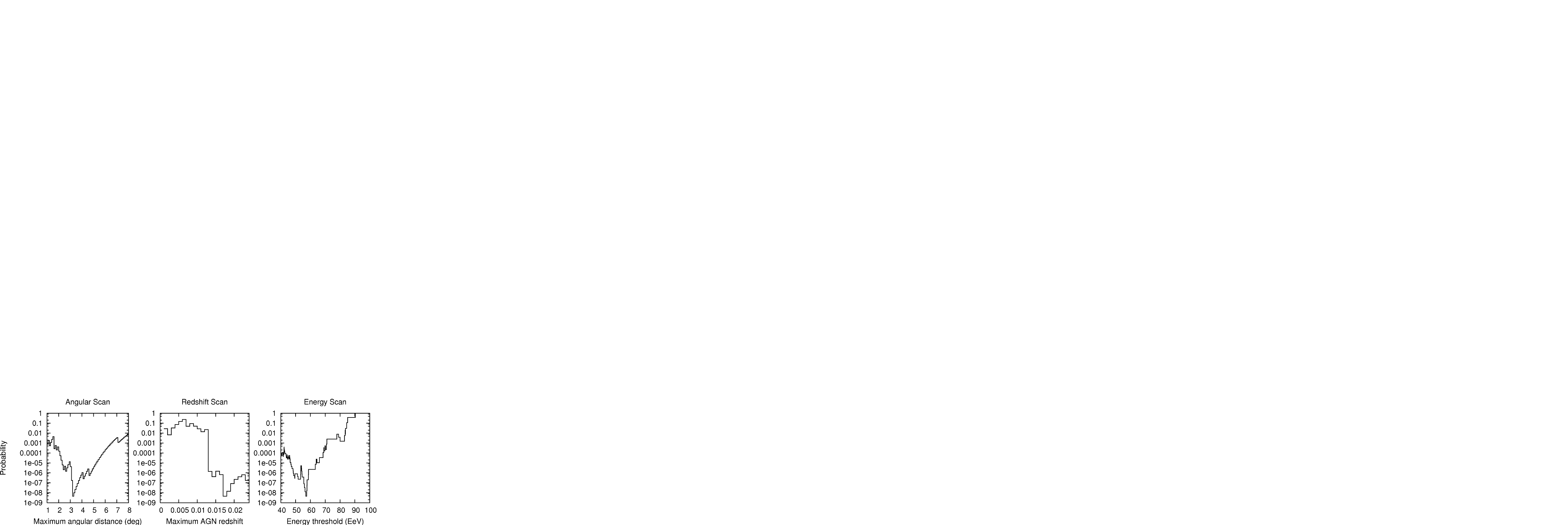}}
\caption{Probability for the null hypothesis (isotropic distribution)  vs. maximum angular 
distance $\psi$ (left), maximum AGN redshift $z_{max}$ (centre), and 
threshold cosmic-ray energy $E_{th}$ (right).
In each case the other two parameters are held fixed at the values that lead to the absolute minimum 
probability ($\psi=3.2^\circ$, $z_{max}= 0.017$, $E_{th}=57$~EeV).}
\label{Pplots}
\end{figure}

In Figure~\ref{Pplots} we show one-dimensional plots of the 
probability $P$
as a function of each of the scan parameters with the other two
held fixed at the values which lead to the absolute minimum
probability.

We note that the energy threshold at which the correlation with nearby
AGN is maximised, i.e., $E_{th}=57$~EeV, matches the energy range at
which the flux measured by the Pierre Auger Observatory is $\sim 50$\%
lower than would be expected from a power law extrapolation of the
spectrum measured at lower energies~\cite{ro07}.  This feature adds
support to the interpretation that the correlation with relatively
nearby sources is evidence for the GZK effect \cite{GZK-1,GZK-2}, as
will be discussed in Section~\ref{s:gzk}.

Relatively small values of $P$ occur for the energy threshold 
$E_{th}\sim60$~EeV for a
range of maximum distances to AGN between 50~Mpc and 100~Mpc and for
angular separations up to $6^\circ$.  For instance,
there is a local minimum with a value $P=8\times 10^{-9}$ very close
to that of the absolute minimum ($P_{min}=4.6\times 10^{-9}$) for the set 
of
parameters $\psi=4.8^\circ$, $z_{max}=0.013$ ($D_{max}=55$~Mpc).  With 
this
set of parameters there are 22 events among the 27 with $E>57$~EeV
that correlate with at least one of the 310 selected AGN, while only
7.4 were expected, on average, to do so by chance if the flux was
isotropic ($p=0.28$).  With limited statistics, the parameters that
minimise the probability $P$ should only be taken as indicative
values of the relevant correlation scales. 

AGN catalogues are likely to be incomplete near the galactic plane,
where extinction from dust in the Milky Way  reduces the 
sensitivity of observations. 
Moreover, cosmic rays that arrive close to the galactic plane are
likely to have been deflected by the magnetic field in the disk more 
than those which arrive with higher galactic latitudes.
These effects could
have some impact upon the estimate of the strength and of the parameters 
that characterise the
correlation. Catalogue incompleteness would weaken the 
measured strength of a true correlation. 

\begin{figure}
\centerline{{\includegraphics[width=.85\textwidth]{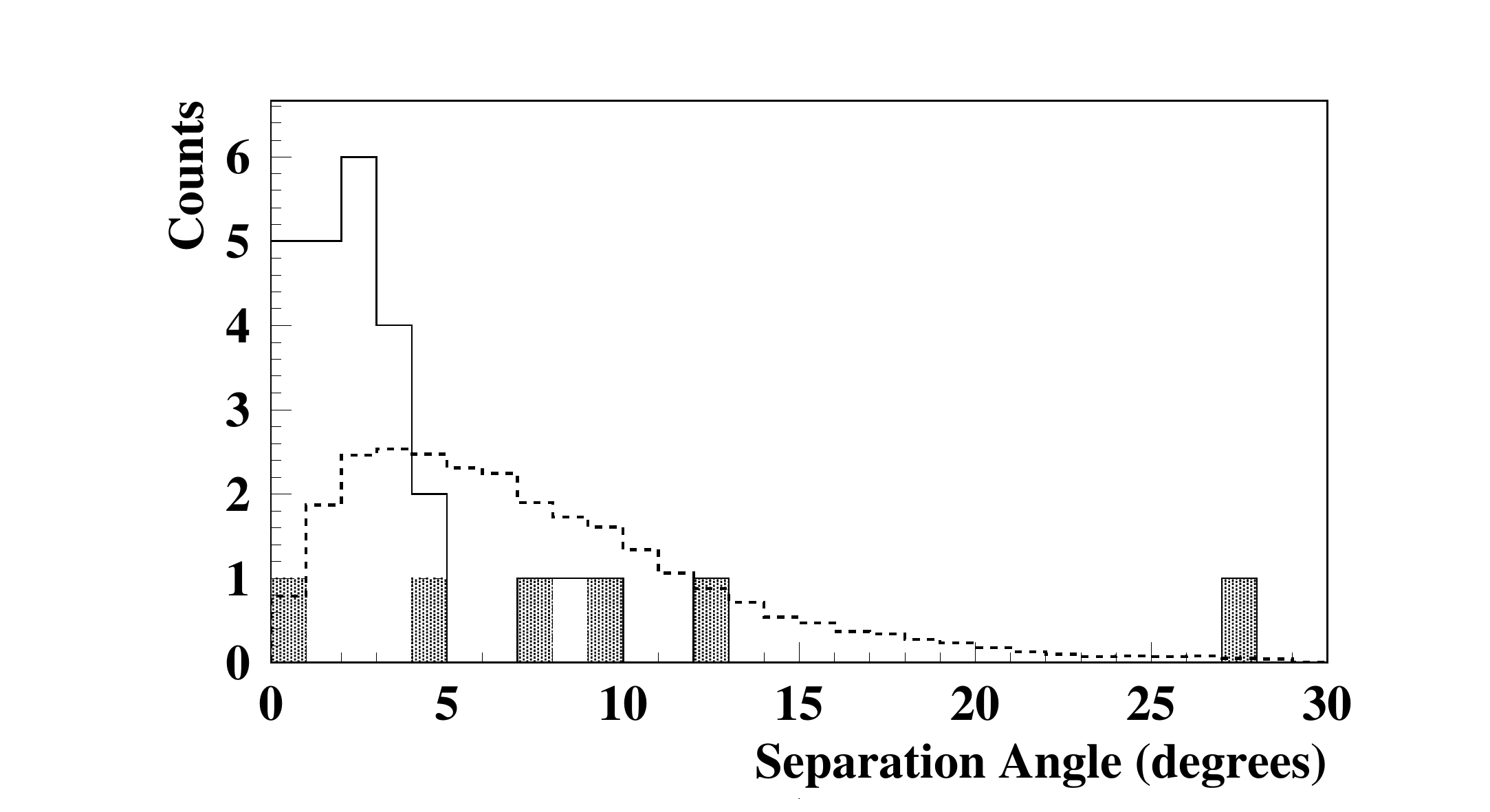}}}
\caption{Distribution of angular separations to the closest AGN
within 71~Mpc. The 6 events with $|b| \le 12^\circ$ have been shaded in grey. The
average expectation for an isotropic flux is shown as the dashed line histogram.}
\label{f:angles}
\end{figure}

In Figure~\ref{f:angles} we plot the distribution of angular 
separations between the arrival directions of the 27 highest-energy events
and the position of the closest AGN with redshift $z\le 0.017$.
On this graph the 6 events with galactic
latitudes $|b|<12^\circ$ have been shaded in grey.
The two distributions are clearly distinct, a likely consequence of the incompleteness of the V-C catalogue at low galactic latitudes. 
The dashed line is, for comparison, the 
distribution
expected, on average, from an isotropic flux modulated by the relative exposure of the Observatory. 

We have performed  a scan limited to events with
galactic latitudes $|b| > 12\degs$.  The minimum probability 
for
the hypothesis of isotropic arrival directions occurs for the same 
parameters 
as without the cut in the galactic plane ($\psi =3.2\degs, z_{max}=0.017, 
E_{th}= 57$~EeV). The cut increases the 
strength of the correlation ($P_{min}=1.7\times 
10^{-10}$).  19 out of 21 arrival directions correlate with AGN 
positions while  5.0 are expected to do so by chance
if the flux were isotropic. In other words, 5 of the 7 events which do
not correlate with AGN positions arrive with galactic latitudes $|b| < 12\degs$. 

\begin{figure}
\includegraphics[width=1.0\textwidth]{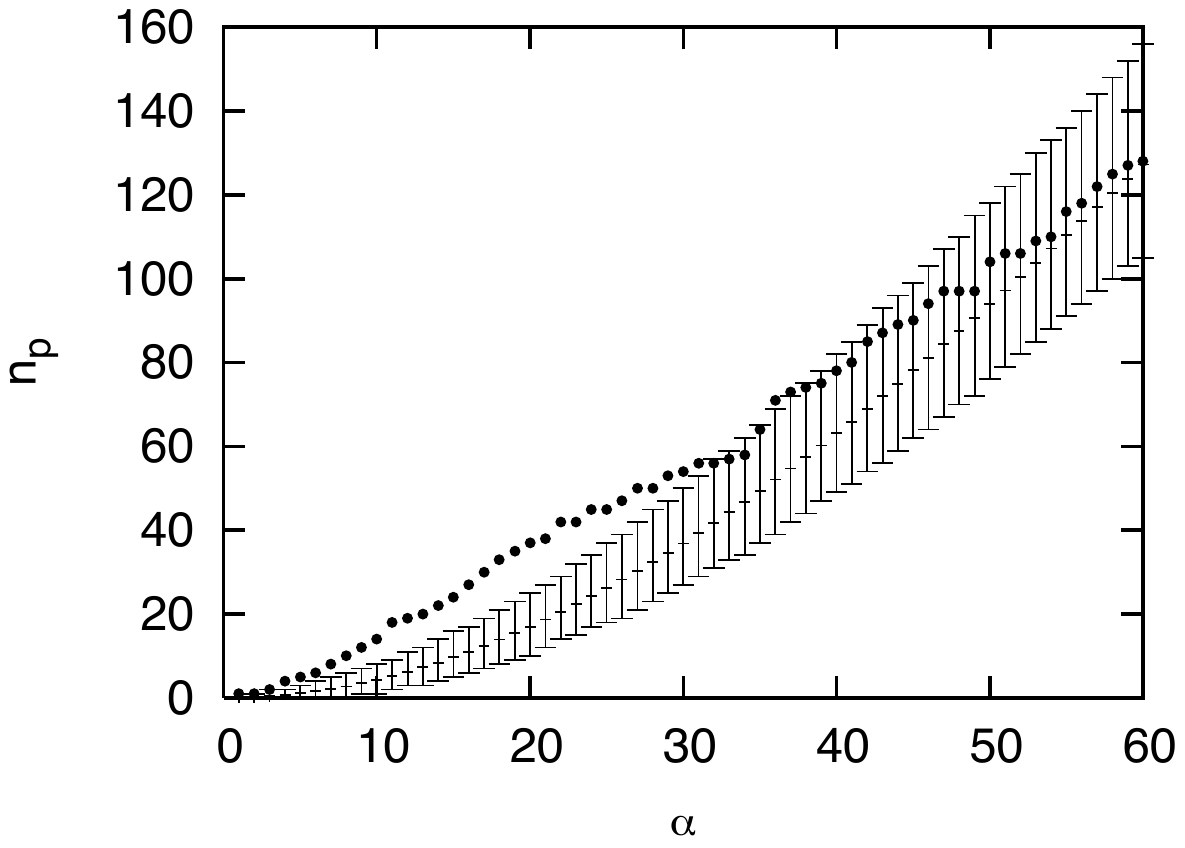}
\caption{Number of pairs
as a function of maximum separation angle $\alpha$ for the 
27 events with 
$E>57$~EeV (points) and average expectation for an isotropic flux. 
The error bars on the isotropic expectations represent the 90\% confidence limit dispersion.}
\label{f:autocorr}
\end{figure}

A distribution of arrival directions of cosmic rays that 
shows evidence of anisotropy by correlation with a  set of
astrophysical objects is also expected to show evidence for
anisotropy by auto-correlation. 
The degree of auto-correlation in the set of the 27 events with $E >
57$~EeV is shown in Figure~\ref{f:autocorr}, where we plotted
the number of event pairs with angular separation smaller than
a given value. Points represent the number of 
pairs in the data. Also
shown are the mean number of pairs expected in simulated isotropic
sets of 27 directions, distributed in proportion to the exposure of
the Observatory. The error bars represent the dispersion of 90\% of the
simulations. Significant departures from
isotropy are seen to occur at intermediate angular scales, between 
$9\degs$
and $22\degs$.  This may be the consequence of a combination of
clustering of events from individual sources in addition to effects of
the non-uniform distribution of the sources themselves
\cite{augerauto}. 

\begin{figure}
\includegraphics[width=1.0\textwidth]{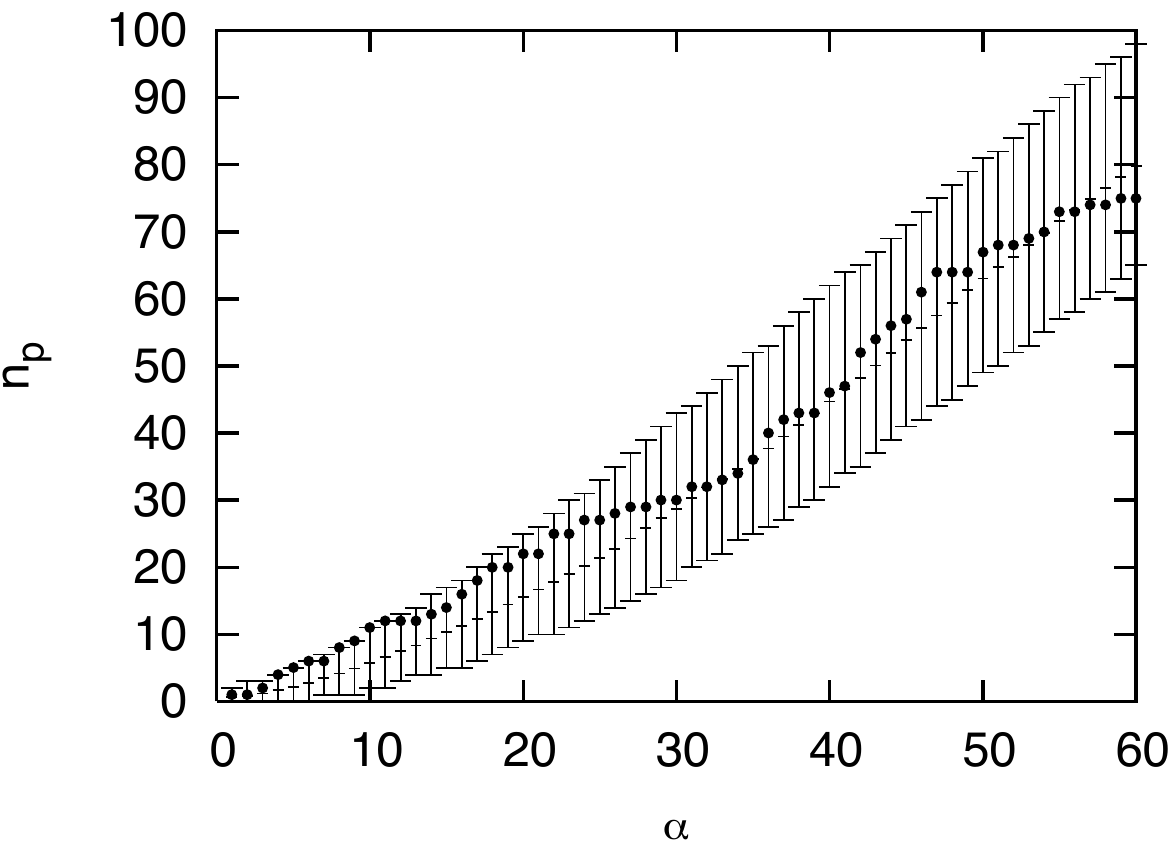}
\caption{ Number of pairs
as a function of maximum separation angle $\alpha$ for the subset of 21 events with 
$E>57$~EeV and $|b| > 12\degs$ (points) and average expectation for 
AGN in the V-C catalogue with $z\le 0.017$ and the same cut in galactic 
latitude. The error bars on the AGN expectations represent the 90\% confidence
limit dispersion.}
\label{f:autocorr-vc}
\end{figure}

To compare the auto-correlation function of the data to that
expected from the AGN distribution in the V-C catalogue we must restrict 
ourselves to the regions where the catalogue is reasonably complete, e.g.,
outside of the galactic plane. In Figure~\ref{f:autocorr-vc} we plot the number
of pairs in the data as a function of the separation angle restricted 
to the 21 events  with $E > 57$~EeV and  galactic latitudes $|b|>12\degs$. 
Also shown is the average distribution expected in sets of 21 directions chosen
at random (in proportion to the relative exposure of the Observatory)
from the positions of AGN in the V-C catalogue with redshift $z\le 0.017$ and 
$|b|>12\degs$. The error bars in the plots indicate the results in 90\% of the simulated sets.
The distribution of pairs in the data are in all cases within those results.

Anisotropy in the distribution of arrival directions can also be
characterised by the multipole moments of an expansion in spherical 
harmonics $Y_{\ell,m}$. With partial coverage of the sky the application
of this method is
not straightforward~\cite{psommers}. A method has been developed~\cite{pb-od} to cope with partial sky coverage that allows a bound to be placed,
with a given confidence level, to the minimum
multipole order necessary to account for the observed
distribution.  The application of this  method to the arrival directions of 
the events with energies above 50 EeV in our data set shows that a 
multipole order $\ell\ge 5$ 
is necessary 
to describe their distribution with 99\% confidence level.
 The arrival directions
of events with energy between 40 EeV and 50 EeV do not show evidence
of anisotropy with this method. This result is consistent with
the sharp decrease of the correlation with AGN positions at lower
energies.  

\section{Discussion}
In the previous sections we have demonstrated the anisotropy of the
highest-energy cosmic rays and have derived a set of parameters that
maximises the correlation with the AGN sample from the 12th edition of
the V-C catalogue. In this section, based on those parameters and the
data set that maximises the correlation, we discuss some possible
implications of the observed signal.

\subsection{Lower limit on the number of sources}
If ultra high energy cosmic rays come from a large number of dim sources,
the number of pairs of events (doublets) coming from one source 
is
expected to be much smaller than the number of singlets. On the other 
hand,
if they come from a small number of bright sources, the ratio of doublets
to singlets is expected to be larger. It is then possible to put a lower
limit on the number of sources based on the ratio of doublets to singlets.
The minimum number of sources, $S$, results for the case in which all
the sources have the same apparent luminosity \cite{dtt00}. If sources are 
steady,
cosmic rays accelerated by one source at different times are statistically
independent and the detection can be considered as a Poisson process.
Then the probability that one source accelerates $n$ particles is given 
by
$P(n) = {\nu}^n e^{-\nu}/n!$, with $\nu$ the mean
number of events expected from one source. The mean number of expected
singlets from $S$ sources is $n_1 = S\times P(1) = S\nu e^{-\nu}$ and that of doublets is
 $n_2= S\times P(2) = S\nu^2 e^{-\nu}/2$. It is thus possible
to estimate the number of sources $S$ as:

\begin{equation}
S \sim \frac{n_1^2}{2 n_2} e^{(2n_2)/n_1}
\end{equation}

Within the 27 highest-energy events there are 6 pairs with separation
smaller than the correlation angular scale of $6^\circ$, while 1.6
 are expected by chance in an isotropic  flux. Taking
$n_2 = 6-1.6=4.4 $ and $n_1 = 27-2n_2=18.2$, we obtain a lower limit for
the number of sources $S \geq 61$.  

Note that this is a bound for mean expectations, but could have 
large
fluctuations with the present small statistics. Also, it was derived 
under the unrealistic assumption of equal flux on Earth for all sources.
Assuming instead equal intrinsic luminosity in cosmic rays the 
mean number of sources becomes larger by a factor of order $n_1/n_2$ 
\cite{dtt00}.
The lower bound could also increase if the sources
had significant clustering of their own on the same angular scale as the
clustering of events.
In either case, this lower limit does not contradict the hypothesis that
nearby AGN are the sources.

\subsection{Signal dependence on energy}\label{V-C}
We have studied the dependence on energy of
the correlation of our data set with the AGN from the V-C
catalogue. In an approach similar to the one developed in~\cite{distrib}
we constructed a smoothed density map from the
V-C catalogue and used it to compute the log-likelihood of any
event sample. We then compared the result for the data to that for simulated 
samples of the same size, either drawn from an isotropic distribution of 
arrival directions or from the smoothed density map itself, in both cases
modulated by the relative acceptance of the Observatory.

From the density map of the V-C catalogue, smoothed on a given angular scale and 
limited to within 100 Mpc, it is possible to calculate  the average log-likelihood of an event sample :
\begin{equation}
LL = \frac{1}{N}\sum_{k=1}^{N} log(\rho_k)
\end{equation}
where the sum runs over the $N$ events of the sample under
consideration and $\rho_k$ is the map density in the 
direction of the event $k$.

\begin{figure}[t]
\begin{center}
\includegraphics[width=1.0\textwidth]{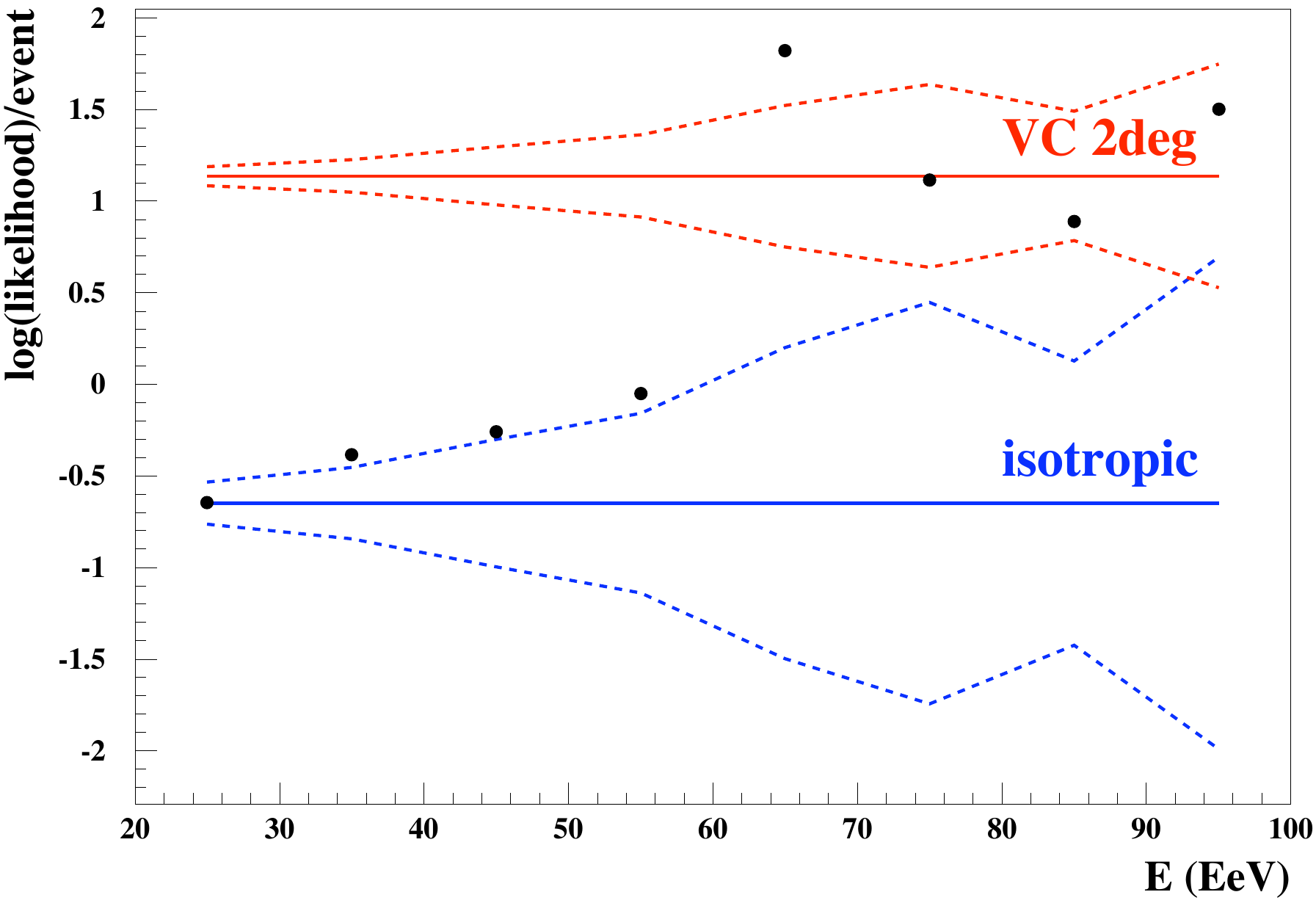}
\end{center}
\caption{\label{LL-lmax15}
Comparison of the average
log-likelihood per event in different energy intervals (calculated using the smoothed
distribution of AGN) between real data and simulated
samples (bottom [blue line] isotropic distribution, top [red line] AGN distribution from V-C catalog). }
\end{figure}

In Figure~\ref{LL-lmax15} we present the values of $LL$ (dots) from our data in 10~EeV energy intervals. We used events with galactic latitudes $|b| > 12^\circ$, and compared them with the average expected from samples of similar size and generated either according to an isotropic distribution (bottom (blue) line) or to the distribution of AGN within 100 Mpc smoothed with Gaussian windows of $2^\circ$ (top (red) line). The dashed lines represent the $1~\sigma$
interval around the mean value, based on the statistics of the real data. The data are compatible with a distribution that follows that of AGN at high energies with an abrupt transition towards an isotropic distribution below 60 EeV.

\subsection{Source identification}
The high degree of correlation that we observe 
can certainly
serve as a strong indication that AGN may well be the sources of
ultra-high energy cosmic rays.  However this result is not yet a
proof.

In particular, we know that the distribution of matter (visible and dark) 
in our local Universe (within 100 Mpc) is strongly
non-uniform, and that AGN are correlated with this non-uniformity.
We are therefore motivated to see if our data provide sufficient
information to determine if the correlation signal is
unambiguously associated with AGN or if they are simply acting as 
tracers of some other set of sources with a similar distribution.

In a specific study we have compared the arrival directions of our highest
energy events with the distribution expected from different source models using a likelihood test.
The test compared isotropic distributions, distributions of galaxies from the 
IRAS PSCz~\cite{IRAS} catalogue and distributions of AGN from the the V-C catalogue, at different angular 
scales and using different horizon depths (see 4.4) .
In addition to confirming that our data are incompatible 
with an isotropic distribution these studies
showed that they are best represented by a small angular smoothing within a  
relatively nearby horizon (100 Mpc) of the AGN from the V-C catalogue~\cite{distrib}.

The local spatial distribution of AGN is correlated with the distribution
of other astronomical objects that are potential sources of cosmic rays, such as rich clusters
of galaxies and star-burst galaxies (which could host a large fraction of gamma ray bursts).
We have tested, with the same scan method that we applied to the
V-C catalogue, whether there is a significant correlation with the Abell clusters of galaxies
(an optical survey) \cite{abell} and with the clusters in the X ray surveys REFLEX \cite{reflex} and
NORAS \cite{NORAS}. We also tested for correlation against a catalogue of starburst galaxies \cite{starburst}. 
We did not find significant correlation in these searches with our present data set.

It is plausible that only a subclass of AGN in the V-C catalogue are
the sources of the highest-energy cosmic rays.
 With our present relatively small data set it is 
difficult to pinpoint distinctive properties of the AGN that are close to
their arrival directions, or to draw firm conclusions about patterns in 
their redshift distribution. It is worth noting, as is clearly visible in 
Figure~\ref{skymap}, the striking alignment of several events close to the
super-galactic plane. Two of the events 
have arrival directions less than 3$^\circ$ away from Centaurus A, one of 
the closest AGN. 

\subsection{The GZK horizon}
\label{s:gzk}
The correlation observed 
is consistent with the hypothesis that 
the highest-energy cosmic rays that arrive on Earth 
are predominantly produced in relatively nearby AGN, within the
distance over which the GZK effect \cite{GZK-1,GZK-2,GZK-3} does not significantly attenuate their flux.

The ``GZK horizon'' may be defined as the distance from the Earth
which contains the sources that produce 90\% of the protons that
arrive with energies above a given threshold.  Under the idealisation
of uniformly distributed sources of equal intrinsic cosmic-ray
luminosity and a conventional spectral index, the horizon computed in
the continuous energy loss approximation is about 90~Mpc for protons
that arrive with energies above 80~EeV and about 200~Mpc for energies
above 60~EeV \cite{ha06}.  Deviations of the horizon scale from the
estimates above are expected, in particular due to local departures of
the sources from uniformity in spatial distribution, intrinsic
luminosity, and spectral features.

The largest departure from isotropic expectations (minimum value of the
probability $P$) in the complete
data set was found to be due to correlation with AGN at a distance smaller than
71~Mpc and 
for cosmic rays with energies above 57~EeV. However, 
 relatively small values of $P$ occur for
this energy threshold
for a range of maximum distances to AGN between 50~Mpc and 100~Mpc.

If these numbers were to be taken at face value, an upward shift in
the energy calibration of $\sim 30$\%, as suggested in some
simulations of the reconstruction of the shower energies \cite{en07},
would lead to a better agreement between the maximum AGN distance
$D_{max}$ that minimises the probability $P$ and the theoretical
expectations based on the idealised GZK attenuation.  However, while
we expect $D_{max}$ to be comparable to the GZK horizon scale, the
relation is not a simple one. An accidental correlation with
foreground AGN different from the actual source may induce some bias
in the value of $D_{max}$ toward smaller maximum source distances.
The $P$ minimisation method has non-uniform sensitivity over the range
of parameters explored. Incompleteness of the V-C catalogue prevents a
reliable exploration beyond approximately 100 Mpc. Furthermore, as
mentioned above, a realistic estimate of the horizon scale depends on
several unknown features. A large local over-density of sources would
reduce its value. The distribution of intensity and spectral features
of the dominant sources also has an effect on the horizon scale.

Regarding the possibility that the cosmic rays injected at the sources
are heavy nuclei, attenuated mainly by photo-disintegration processes,
one may note that nuclei of the iron group have horizons only slightly
smaller than the proton horizons, but intermediate mass nuclei
($A\simeq 20$--40) have significantly smaller horizons (e.g., the
horizon for a threshold energy of 60~EeV is about 60~Mpc for $^{28}$Si
nuclei \cite{ha06}). The smaller horizon for decreasing nuclear mass
is due to the corresponding decrease in the threshold required to
excite the giant-dipole resonance for photo-disintegration.

\subsection{Effect of the magnetic fields}
\label{s:angular}

A cosmic ray with charge $Ze$ that travels a 
distance $D$ in a regular magnetic field B is deflected by 
an angle $\delta$ given by
\begin{equation}
\delta \simeq 2.7^{\circ}\frac{60\textrm{ EeV}}{E/Z}\left|\int\limits_{0}^{D}
\left(\frac{\textrm{d}\mathbf{x}}{\textrm{ kpc}} \times 
\frac{\mathbf{B}}{3\;\mu \textrm{G}}\right) 
\right|
\end{equation}

If  the regular galactic magnetic field has a strength of  a few $\mu$G with a coherence scale 
of order $\sim$ 1 kpc, as in some models~\cite{stanev},  the deflection  is 
expected to be a few degrees for protons with $E > 60$ EeV.  In such models,
the angular scale of the correlation we observed is consistent with the  size 
of the deflections expected to be imprinted upon protons by the galactic magnetic field.

\begin{figure}[t]
  \begin{minipage}{\textwidth} 
    \begin{minipage}[t]{0.5\textwidth} 
      \begin{center}
        \includegraphics[width=0.95\textwidth]{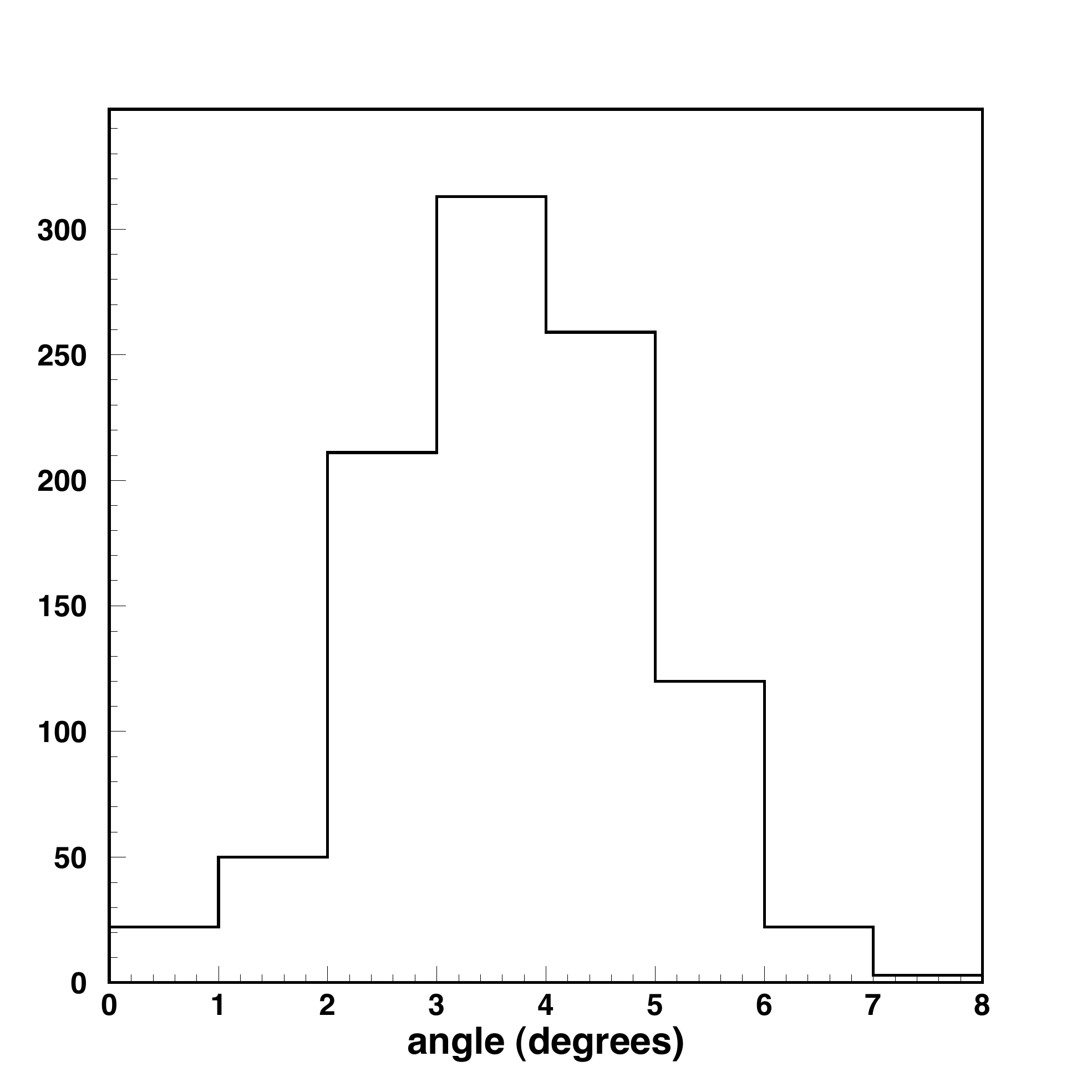}
      \end{center} 
    \end{minipage} 
    \begin{minipage}[t]{0.5\textwidth} 
      \begin{center} 
        \includegraphics[width=0.95\textwidth]{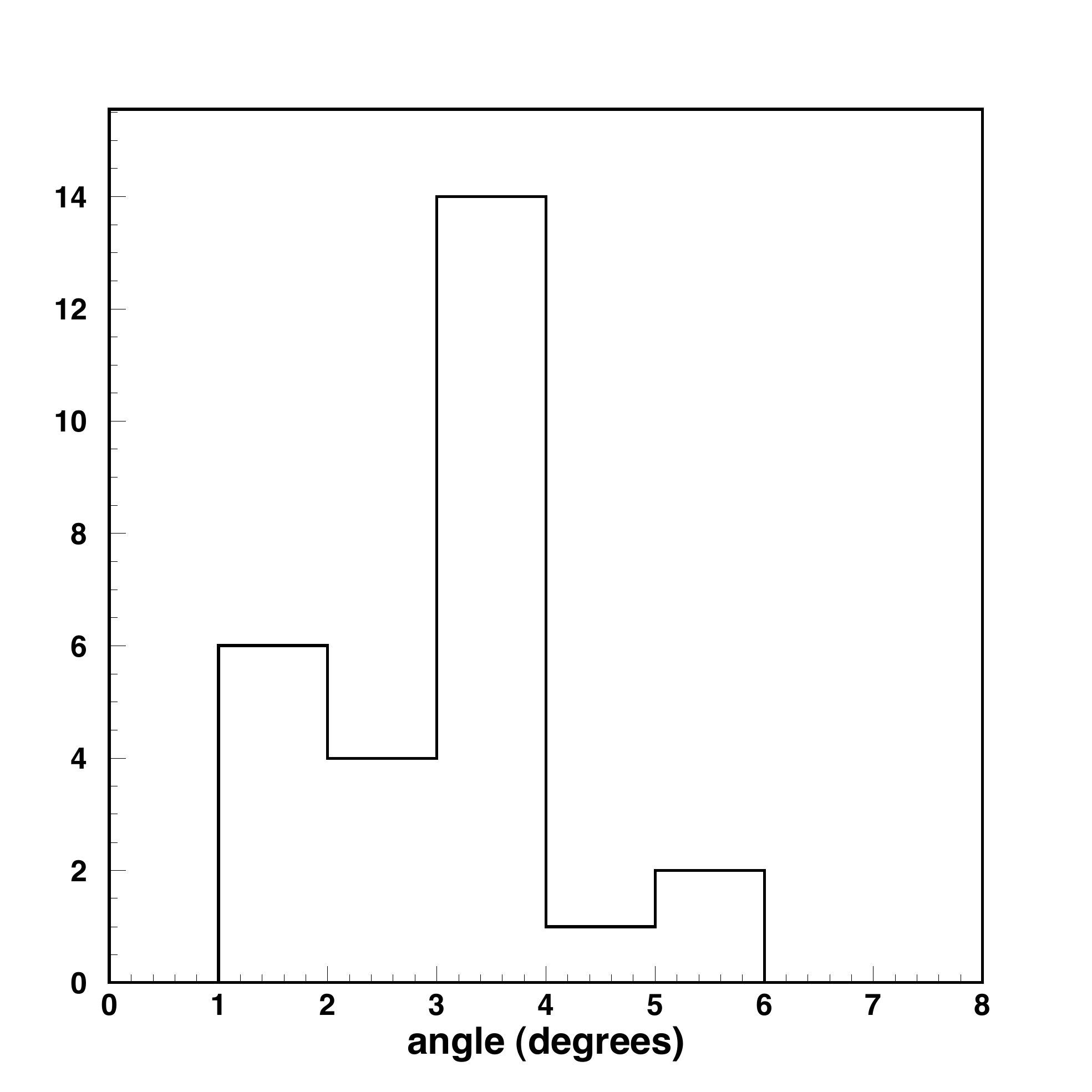}
      \end{center} 
    \end{minipage} 
  \end{minipage}
\caption{Distribution of the deflections for protons in the BSS-S model of 
the galactic magnetic field. Left panel: 1000 directions drawn from an isotropic 
flux in proportion to the exposure of the Observatory, for $E=$ 60 EeV. Right 
panel: deflections of the 27 arrival directions of the observed events with 
$E>$57~EeV
.}
\label{histodevia}
\end{figure}

The precise amount of the deflection is very dependent on each specific
arrival direction. We have evaluated numerically the deflections imparted 
in a conventional regular galactic magnetic field model \cite{stanev} 
for sets of arrival directions uniformly distributed according  to the 
Pierre Auger Observatory relative exposure. Anti-particles were backtracked in the galactic regular 
magnetic field to a distance of 20 kpc away from the Galactic Centre 
(where the field strength is already very small). At this point the 
angle between the initial (as measured on the Earth) and final velocity
vectors was calculated. The result is shown in the left panel of Figure 
 \ref{histodevia} in the special case of the BSS-S model\footnote{In fact, 
we have smoothed the original BSS-S model of \cite{stanev} as 
described in \cite{ha99} in order to avoid the discontinuities 
present in the original model.} (without $B_z$ component) for $E = 60$ EeV protons. 
The deflections scale approximately 
as $Z \times (60 \textrm{ EeV}/E)$ for other energies and electric charges
(the scaling is rigorous only for very small deflections and a uniform 
field). In the right panel of the figure we show the distribution of 
deflections for protons in the case of the 27 arrival directions of the 
events with $E > 57$~EeV, as computed for each using its reconstructed energy. 

Models of the regular component of the galactic magnetic field  \cite{Breview} 
outline its basic features, but cannot be expected to provide a 
complete picture nor a realistic value for every direction.  It is, for example,
possible to do the exercise of ``correcting'' the observed  arrival  directions 
to undo the deviation imparted by the galactic magnetic field, but 
current models are not expected to be accurate enough 
to allow us to draw reliable conclusions from such analyses.
Nonetheless, the results shown in Figure~\ref{histodevia} provide a reasonable estimate of the typical 
deflections to be expected. They are consistent with the angular 
scale of the observed correlation with AGN. 
Therefore, if the BSS-S model is a fair representation of the general features of the 
regular galactic magnetic field, then the correlation observed in the data would be
unlikely if the primary composition of the cosmic rays reaching us were much
heavier than protons. Note that this does not preclude the possibility that
the source emits heavy nuclei, which could disintegrate along their journey,
so that the lighter fragments are those deflected by the galactic magnetic field.

It will be possible in the future to compare various models for the
galactic magnetic field with the pattern of orientation and size of
the deviation between the observed arrival directions of the events
and potential AGN sources. However, our present data set is not
large enough to perform such an analysis reliably.

The angular scale of the observed correlation also implies that intergalactic 
magnetic fields along the line of sight to the sources do not in general deviate cosmic-ray 
trajectories by much more than a few degrees. The root-mean-square deflection imprinted 
upon the trajectories of cosmic rays with charge $Ze$ as they travel a distance $D$ 
in a turbulent magnetic field with coherence length $L_c$ is
\begin{equation}
\delta_{rms} \approx 4^{\circ}\frac{60\textrm{ EeV}}{E/Z}
\frac{B_{rms}}{10^{-9}\textrm{G}}\sqrt{\frac{D}{100\textrm{ Mpc}}}
\sqrt{\frac{L_{c}}{1\textrm{ Mpc}}}
\end{equation}

There is no measurement of the intergalactic magnetic fields
except at the centres of rich galaxy clusters.  Numerical simulations of 
those fields give a wide range of 
possible deflections from negligible \cite{Dolag:2004kp} to very large 
\cite{Sigl:2004yk}. 
The correlation observed can be used to constrain models of
turbulent intergalactic magnetic fields, which must be  such that in most
directions  
$B_{rms}\sqrt{L_{c}} 
\le 10^{-9} \textrm{G}\sqrt{\textrm{Mpc}}$ within our 
``GZK horizon''.

Finally, there are possible biases in the determination of the relevant 
angular scale of the deflection. The active galaxy closest to the arrival direction 
of a cosmic ray is not necessarily the source responsible for it. This  could lead 
to an underestimate of the deflection involved. In principle it could  
also happen, due to catalogue incompleteness or because the source is 
something else, that an AGN found to correlate with a 
particular event is further away than the actual source, thus
overestimating the deflection angle.

\subsection{The acceleration sites}
Acceleration sites in the active galaxies that correlate with events
above 57~EeV are promising candidate sources of high energy cosmic
rays, but other possible sites cannot be ruled out with the present
limited statistics.  (For a recent summary of proposed acceleration
sites see, e.g., \cite{Stanev04}.) The observed correlation shows that
ultra-high energy cosmic ray sources are extra-galactic with an
angular distribution similar to that of AGN within $\sim$ 71 Mpc and
that the primaries are most likely protons that suffer losses due to
interactions with the cosmic background radiation. These results rule
out models for the origin of cosmic rays that place observed sources
predominantly in our Galaxy, such as galactic compact objects (young
neutron stars \cite{ynsw}, pulsars \cite{pulsar}, and black holes),
and gamma-ray bursts \cite{GGRB}. Models where sources are located in
the galactic halo are also ruled out, such as the decay of super-heavy
dark matter particles \cite{SHDM-1,SHDM-2,SHDM-3}, which are already
highly constrained by the Auger limit on the fraction of photon
primaries at high energies \cite{photonlimit,TDlimit}. Top-down models based on
topological defects \cite{TD-1,TD-2,TD-3} need to have a spatial
distribution consistent with the local matter distribution to avoid
being excluded as significant sources of cosmic rays. Such models are
also constrained by the photon and neutrino limits \cite{TDlimit}.

The large-scale structure distribution of matter, which is traced by normal 
galaxies, has a similar spatial distribution to the local AGN. Therefore, 
acceleration sites in galaxies with inactive nuclei cannot be excluded at the 
present, including those based on extra-galactic compact objects \cite{EGco}, 
quasar remnants \cite{QR}, galactic winds in star-bursts galaxies 
\cite{starb}, and gamma-ray bursts \cite{GRBs,GRBsA}. In contrast, 
acceleration 
models in massive clusters of galaxies, such as cluster accretion 
shocks \cite{CluS,CluSA}, are challenged by the observed correlation. 
Massive 
clusters are rare within 100 Mpc when compared to the number of observed 
events and there is a paucity of events from the direction of Virgo, the 
nearest sizeable cluster of galaxies.

AGN have long been suggested as likely accelerators of cosmic rays 
\cite{Ginzburg,Hillas,AGNref}.
The case for active galaxies as likely 
sources is based on the power available from the central black hole. 
AGN are powered by the accretion of matter onto a super-massive black hole 
(with masses in the range $10^6$ to $10^8$ M$_{\odot}$) at the centre of the 
galaxy. A number of mechanisms have been proposed that utilise different 
regions and properties of this system to accelerate cosmic rays to ultra-high 
energies. Acceleration based on the central regions face the challenge of 
energy losses in the radiation field that surrounds the central black hole 
and accretion disk. Alternative acceleration sites include jets \cite{AGNjet,AGNjetA} 
and radio lobes \cite{AGNrlob} that are associated with the most luminous 
AGN.

AGN with prominent radio lobes are rare and do not follow the observed spatial 
distribution of the observed correlated AGN.  The one exception is Cen A, at only 
3.4 Mpc~\cite{CenA-dist},
 which has been proposed as a site for cosmic-ray acceleration \cite{CenA-0}. 
 It displays jets, radio lobes which extend
over a scale of about $10^\circ$ along the super-galactic plane, and a variable 
compact radio nucleus. 
Two events correlate with the nucleus position while several lie in the vicinity of the radio lobe
extension along the super galactic plane (see Figure~\ref{skymap}).
 The most prominent radio galaxy in our GZK neighbourhood is M87, which does not 
correlate with any observed event above 57~EeV thus far but the coverage of the southern
Auger Observatory is almost a factor of 3 lower in this direction of the sky than in the direction of Cen A.
Of the remaining 18  correlating events, 15 have Seyfert galaxies as the closest AGN in angular 
separation.

A significant increase in ultra-high energy cosmic-ray statistics
combined with searches for counterparts in a multi-wavelength and
multi-messenger campaign should improve our ability to distinguish if
AGN are the sources of cosmic rays or tracers of the sources. If
future data select AGN as hosts of cosmic-ray accelerators, the type
of AGN selected, together with spectral and composition information,
should help distinguish between proposed AGN acceleration mechanisms.

\section{Conclusions}
Anisotropy has been established with more than 99\% confidence level 
in the arrival directions of 
events with energy above $\sim$ 60 EeV detected by the  Pierre Auger 
Observatory. These events correlate over angular scales of less than $6^\circ$ 
with the directions towards nearby ($D < 100$~Mpc) AGN. 

The observed correlation demonstrates the extra-galactic 
origin of the highest-energy cosmic rays. It is consistent 
with the hypothesis that cosmic rays with energies above 
$\sim$~60~EeV are predominantly
protons that come from AGN within our ``GZK horizon".  This provides
evidence that the observed steepening of the cosmic-ray spectrum at
the highest energies is due to the ``GZK effect", and not to acceleration
limits at the sources.

It is  possible that the sources are other than AGN, as long as their local distribution is
sufficiently correlated with them. 
Unequivocal identification of the sources requires a larger data set, such as 
the
Pierre Auger Observatory will gather in a few years.
In particular, one could use the fact that 
angular departures of the events
from an individual source due to magnetic deflections should decrease in
inverse proportion to the energy of the cosmic ray. The observation of 
such
angle/energy correlation in clusters of events could be exploited to
locate the source position unambiguously with high accuracy.
This could also provide at
the same time valuable and unique information about the magnetic fields 
along
the line of sight.  

We have shown that astronomy of charged particles 
is indeed feasible at the highest energies and that in the next few years we can hope for 
unambiguous identification of sources of cosmic rays. 

\section*{Acknowledgments}
The successful installation and commissioning of the Pierre Auger Observatory
would not have been possible without the strong commitment and effort
from the technical and administrative staff in Malarg\"ue.

We are very grateful to the following agencies and organizations for financial support: 
Comisi\'on Nacional de Energ\'ia At\'omica, Fundaci\'on Antorchas,
Gobierno De La Provincia de Mendoza, Municipalidad de Malarg\"ue,
NDM Holdings and Valle Las Le\~nas, in gratitude for their continuing
cooperation over land access, Argentina; the Australian Research Council;
Conselho Nacional de Desenvolvimento Cient\'ifico e Tecnol\'ogico (CNPq),
Financiadora de Estudos e Projetos (FINEP),
Funda\c{c}\~ao de Amparo \`a Pesquisa do Estado de Rio de Janeiro (FAPERJ),
Funda\c{c}\~ao de Amparo \`a Pesquisa do Estado de S\~ao Paulo (FAPESP),
Minist\'erio de Ci\^{e}ncia e Tecnologia (MCT), Brazil;
Ministry of Education, Youth and Sports of the Czech Republic;
Centre de Calcul IN2P3/CNRS, Centre National de la Recherche Scientifique (CNRS),
Conseil R\'egional Ile-de-France,
D\'epartement  Physique Nucl\'eaire et Corpusculaire (PNC-IN2P3/CNRS),
D\'epartement Sciences de l'Univers (SDU-INSU/CNRS), France;
Bundesministerium f\"ur Bildung und Forschung (BMBF),
Deutsche Forschungsgemeinschaft (DFG),
Finanzministerium Baden-W\"urttemberg,
Helmholtz-Gemeinschaft Deutscher Forschungszentren (HGF),
Ministerium f\"ur Wissenschaft und Forschung, Nordrhein-Westfalen,
Ministerium f\"ur Wissenschaft, Forschung und Kunst, Baden-W\"urttemberg,
Germany; Istituto Nazionale di Fisica Nucleare (INFN),
Ministero dell'Istruzione, dell'Universit\`a e della Ricerca (MIUR), Italy;
Consejo Nacional de Ciencia y Tecnolog\'ia (CONACYT), Mexico;
Ministerie van Onderwijs, Cultuur en Wetenschap,
Nederlandse Organisatie voor Wetenschappelijk Onderzoek (NWO),
Stichting voor Fundamenteel Onderzoek der Materie (FOM), Netherlands;
Ministry of Science and Higher Education,
Grant Nos. 1 P03 D 014 30, N202 090 31/0623, and PAP/218/2006, Poland;
Funda\c{c}\~ao para a Ci\^{e}ncia e a Tecnologia, Portugal;
Ministry for Higher Education, Science, and Technology,
Slovenian Research Agency, Slovenia;
Comunidad de Madrid, Consejer\'ia de Educaci\'on de la Comunidad de Castilla
La Mancha, FEDER funds, Ministerio de Educaci\'on y Ciencia,
Xunta de Galicia, Spain;
Science and Technology Facilities Council, United Kingdom;
Department of Energy, Contract No. DE-AC02-07CH11359,
National Science Foundation, Grant No. 0450696,
The Grainger Foundation USA; ALFA-EC / HELEN,
European Union 6th Framework Program,
Grant No. MEIF-CT-2005-025057, and UNESCO.

\clearpage
\appendix

\section{Event List}\label{EventList}
Here we list the 27 events recorded from 1 January 2004 until 31
August 2007 with energy in excess of 57 EeV.  We have indicated the
date of observation (year and Julian day), the zenith angle, the
shower size at 1000~m from the core $S(1000)$, the energy in EeV, the
equatorial coordinates (RA, Dec) and the galactic coordinates
(Longitude, Latitude).  Events that correlate within $3.2\degs$ of AGN
with redshift $z\le 0.017$ are marked with a star. The dashed
horizontal line indicates the beginning of the prescribed test of
section 2.

The quoted energy is derived from a calibration procedure where 
the shower size is compared to the energy measured by the FD. 
This energy calibration, based on the sample of hybrid events analysed 
at the time of the prescription, was used for the whole data set for consistency. 
The smaller uncertainty on the energy calibration curve expected from the increased
statistics of hybrid events, as well as improvements in the systematic
uncertainty of the FD energy scale, may lead to revised energies in future
publications. Thus, we also include the shower size S(1000) at 1000 m
from the reconstructed core,  as it is the shower parameter that is directly measured from
the individual SD signals in the event. This parameter is almost
independent of the shower lateral distribution function used in the
reconstruction procedure (within 10\%).
The uncertainty in $S$ resulting from the adjustment of the shower size, the conversion to a reference angle, the fluctuation from shower to shower and the calibration curve amounts  to about 18\%.
The absolute energy scale is given by the fluorescence measurements and has a systematic uncertainty of 22\%~\cite{bruce}.

\begin{table}[htbp]
\begin{center}
\begin{tabular}{ccccccccc}\hline
Year & Julian day & $\theta$ & $S(1000)$ & E (EeV) & RA & Dec & Longitude & Latitude  \\ \hline\hline
2004 & 125 & 47.7 &  252 & 70 & 267.1$^\circ$ & -11.4$^\circ$ &   15.4$^\circ$&  8.4$^\circ$ \\
2004 & 142 & 59.2 & 212 & 84 & 199.7$^\circ$ & -34.9$^\circ$ &   -50.8$^\circ$&  27.6$^\circ$ * \\
2004 & 282 &  26.5 & 328 & 66 & 208.0$^\circ$ & -60.3$^\circ$ &  -49.6$^\circ$&  1.7$^\circ$ *\\
2004 & 339 &  44.7 & 316 & 83 & 268.5$^\circ$ & -61.0$^\circ$ &  -27.7$^\circ$&-17.0$^\circ$ *\\
2004 &  343 & 23.4 & 323  & 63 & 224.5$^\circ$& -44.2$^\circ$ & -34.4$^\circ$ & 13.0$^\circ$ *\\
2005 & 54 & 35.0 & 373 & 84 &  17.4$^\circ$ & -37.9$^\circ$  & -75.6$^\circ$&-78.6$^\circ$ *\\
2005 & 63 & 54.5 & 214 &71& 331.2$^\circ$ &  -1.2$^\circ$ &   58.8$^\circ$&-42.4$^\circ$ \\
2005 & 81 & 17.2 & 308& 58 & 199.1$^\circ$ & -48.6$^\circ$ &  -52.8$^\circ$& 14.1$^\circ$ *\\
2005 & 295 & 15.4 & 311& 57 & 332.9$^\circ$ & -38.2$^\circ$ &    4.2$^\circ$&-54.9$^\circ$ *\\
2005 & 306 &  40.1 & 248 & 59 & 315.3$^\circ$ &  -0.3$^\circ$ &   48.8$^\circ$&-28.7$^\circ$ *\\
2005 & 306 & 14.2 & 445 & 84 & 114.6$^\circ$ & -43.1$^\circ$ &  -103.7$^\circ$&-10.3$^\circ$ \\
2006 & 35 &  30.8 & 398 & 85 &  53.6$^\circ$ &  -7.8$^\circ$  & -165.9$^\circ$&-46.9$^\circ$ *\\
2006 & 55 & 37.9 & 255 & 59 & 267.7$^\circ$ & -60.7$^\circ$ &  -27.6$^\circ$&-16.5$^\circ$ *\\
2006 & 81 &  34.0 & 357 & 79 & 201.1$^\circ$ & -55.3$^\circ$ &  -52.3$^\circ$&  7.3$^\circ$ \\
\hdashline
2006 & 185 &  59.1 & 211 & 83 & 350.0$^\circ$ &   9.6$^\circ$ &   88.8$^\circ$&-47.1$^\circ$ *\\
2006 & 296 & 54.0 & 208 & 69 &  52.8$^\circ$ &  -4.5$^\circ$  & -170.6$^\circ$&-45.7$^\circ$ *\\
2006 & 299 &  26.0 & 344 & 69 & 200.9$^\circ$ & -45.3$^\circ$ &  -51.2$^\circ$& 17.2$^\circ$ *\\
2007 & 13 &14.3 &762& 148& 192.7$^\circ$ & -21.0$^\circ$ &  -57.2$^\circ$& 41.8$^\circ$ \\
2007 & 51 & 39.2 & 247 & 58 & 331.7$^\circ$ &   2.9$^\circ$ &   63.5$^\circ$&-40.2$^\circ$ *\\
2007 & 69 &  30.4& 332 & 70 & 200.2$^\circ$ & -43.4$^\circ$ &  -51.4$^\circ$& 19.2$^\circ$ **\\
2007 & 84 & 17.3 & 340 & 64 & 143.2$^\circ$ & -18.3$^\circ$ &  -109.4$^\circ$& 23.8$^\circ$ *\\
2007 & 145 & 23.9 & 392 & 78 &  47.7$^\circ$ & -12.8$^\circ$ &  -163.8$^\circ$&-54.4$^\circ$ *\\
2007 & 186 &  44.8 & 248 & 64 & 219.3$^\circ$ & -53.8$^\circ$ &  -41.7$^\circ$&  5.9$^\circ$ \\
2007 & 193 & 18.0 & 469 &  90 & 325.5$^\circ$ & -33.5$^\circ$ &   12.1$^\circ$&-49.0$^\circ$ *\\
2007 & 221 & 35.3 & 318 & 71  & 212.7$^\circ$ & -3.3$^\circ$ &   -21.8$^\circ$&54.1$^\circ$ *\\
2007 & 234 &  33.2 & 365 & 80& 185.4$^\circ$ & -27.9$^\circ$ &  -65.1$^\circ$&34.5$^\circ$ \\
2007 & 235 &  42.6&  276 & 69& 105.9$^\circ$ & -22.9$^\circ$ &  -125.2$^\circ$&-7.7$^\circ$ \\\hline
\end{tabular}
\end{center}
\label{EventTable}
\end{table}

Note that the energies and arrival directions given in this list
correspond to the analysis of the full data set which used a slightly
different reconstruction package than the one used for the original
scan and the prescribed test of section 2. In particular improvements
made in the SD tank calibration have very slightly modified the energy
and arrival directions.

If one were to apply the prescribed parameters to this particular
reconstruction, the prescription would have been fulfilled earlier
with the event 2007-069 (10 March 2007, marked with a double star in
the table), with 5 events in correlation out of 6 above 56~EeV.

\end{document}